\documentclass[useAMS,usenatbib]{mn2e}

\usepackage{amssymb,amsmath,graphicx}
\usepackage{footnote}
\usepackage{lscape}
\usepackage{etex}
\usepackage{upgreek}
\usepackage{color}
\usepackage{xcolor}

\newcommand{\AIPS}{{$\cal AIPS\/$}}
\usepackage{caption}
\usepackage[justification=centering]{caption}
\usepackage{ulem}

\def\lfir{L_{\rm FIR}}
\def\ltir{L_{\rm TIR}}

\title[The Far-Infrared-Radio Correlation in MS0451-03]{ {\bf  The Far-Infrared-Radio Correlation in MS0451-03}}
\author[S. M. Randriamampandry et al.]{S. M. Randriamampandry,$^{1,2}$\thanks{E-mail:
solohery@saao.ac.za} S. M. Crawford,$^{1}$ C. M. Cress,$^{2,4}$  K. M. Hess,$^{3}$ \newauthor M. Vaccari,$^{2}$ E. M. Wilcots,$^{5}$ M. A. Bershady$^{5}$ and G. D. Wirth$^{6}$ \\ 
$^1$South African Astronomical Observatory,  P.O. Box 9, Observatory 7935, South Africa \\ 
$^2$Department of Physics, University of the Western Cape, Private Bag X17, Bellville 7535, South Africa \\
$^3$Astrophysics, Cosmology and Gravity Centre (ACGC), Astronomy Department, University of Cape Town, \\ Private Bag X3, 7701 Rondebosch, South Africa \\
$^4$Centre for High Performance Computing, 15 Lower Hope Street, Rosebank, Cape Town, South Africa \\
$^5$University of Wisconsin-Madison, 475 North Charter Street Madison, WI 53706, USA \\
$^6$W.M. Keck Observatory, 65-1120 Mamalahoa Hwy, Kamuela HI 96743, USA}

\begin{document}

\date{Accepted 2014 November 13.  Received 2014 November 12; in original form 2013 September 16}


\pagerange{\pageref{firstpage}--\pageref{lastpage}} \pubyear{2014}

\maketitle

\label{firstpage}

\begin{abstract}
We present a multi-wavelength analysis of star-forming  galaxies in the massive cluster MS0451.6-0305 at z $\sim$ 0.54 to shed new light on the evolution of the far-infrared-radio relationship in distant rich clusters. We have derived total infrared luminosities for a spectroscopically confirmed sample of cluster and field galaxies through an empirical relation based on $Spitzer$ MIPS 24 $\upmu$m photometry. The radio flux densities were measured from deep Very Large Array 1.4 GHz radio continuum observations. We find the ratio of far-infrared to radio luminosity for galaxies in an intermediate redshift cluster to be $q_{\rm FIR}$ = 1.80$\pm$0.15 with a dispersion of 0.53.  Due to the large intrinsic dispersion, we do not find any observable change in this value with either redshift or environment. However, a higher percentage of galaxies in this cluster show an excess in their radio fluxes when compared to low redshift clusters ($27^{+23}_{-13}\%$ to $11\%$), suggestive of a cluster enhancement of radio-excess sources at this earlier epoch. In addition, the far-infrared-radio relationship for blue galaxies, where $q_{\rm FIR}$ = 2.01$\pm$0.14 with a dispersion of 0.35, is consistent with the predicted value from the field relationship, although these results are based on a sample from a single cluster. 
\end{abstract}


\section{Introduction} \label{sec:Introduction}
Radio continuum emission from normal star-forming galaxies can be a powerful tracer of recent star formation activity \citep{1992ARA&A..30..575C}. The radio luminosities at 1.4 GHz are tightly correlated with the far-infrared luminosities for various galaxy types \citep[e.g.][]{1971A&A....15..110V,1985ApJ...298L...7H,1991ApJ...376...95C} over a wide range of redshift \cite[e.g.][]{2002A&A...384L..19G,2004ApJS..154..147A,2010ApJS..186..341S,2010MNRAS.409...92J,2010MNRAS.402..245I}. 

The correlation is believed to be driven by the internal star formation rate. Radio emission from these galaxies are predominantly produced from the synchrotron emission of cosmic-ray electrons accelerated in supernova shocks. The infrared emission is due to ultraviolet light from young massive stars that is absorbed and re-radiated by dust \citep{1992ARA&A..30..575C}. However, it is still unclear what maintains this strong correlation seen over such a wide range of galaxies \citep{2009ApJ...706..482M}. 

The relationship shows lower far-infrared to radio luminosity ratios in galaxy clusters than that found in the field \citep{1995AJ....109.1582A,2004ApJ...600..695R} with much of the variation coming from a subset of objects with large deviations from the relationship \citep{2001AJ....121.1903M}. A number of different processes drive the evolution of galaxies in clusters such as gravitational interactions and ram pressure \cite[see e.g.,][for a review]{2006PASP..118..517B}, and result in transforming blue, star-forming galaxies into the ubiquitous red, quiescent galaxies that dominate cluster populations today. These physical processes have been invoked to explain the differences seen in the far-infrared-radio relationship as measured between the cluster and field \citep{2009ApJ...694.1435M}. 

However, little work has been done at higher redshifts where we see an increase of star-forming  galaxies \citep{1984ApJ...285..426B}, transitional galaxies like E+A \citep{1996MNRAS.279....1B}, and AGN \citep{2009ApJ...701...66M} in galaxy clusters. Despite studies looking at the multi-wavelength properties of galaxies in distant clusters \cite[e.g.][]{2002MNRAS.330...17B,2008ApJ...685L.113S}, no systematic study has been made of the far-infrared-radio relationship in these clusters, and so, the present work is an unique opportunity to explore part of parameter space in {\it redshift} and {\it environment} that has not previously been probed. 
 
This work aims to measure the far-infrared-radio relation in the massive galaxy cluster MS0451.6-0305 (hereafter, MS0451-03) to test how this relationship changes at intermediate redshift between the field and a high-density cluster environment. The properties of the cluster are summarised as follows: MS0451-03 is a massive (M$_{200}$ $\sim$ 3$\times$10$^{15}$ M$_{\odot}$), X-ray luminous (L$_{\rm x}^{\rm bol}$=4$\times$10$^{45}$ erg s$^{-1}$), and large (R$_{200}$ $\sim$ 2.5 Mpc) cluster at $z$=0.538 \citep[see][Table 1]{2009ApJ...690.1158C}.

The paper is organised as follows. Section \ref{sec:Observations} presents the VLA 1.4 GHz radio continuum observations and provides our radio data reduction and analysis. Section \ref{Spitzer Observation and Photometry} presents the $Spitzer$ MIPS observations along with the infrared (IR) photometry. Section \ref{sec:Sample Selection} and Section \ref{sec:Method} describe our sample selection procedure and methodology. Section \ref{sec:Results} presents our results. Finally, Section \ref{sec:Discussions} and Section \ref{sec:Conclusion} discuss and summarise our findings and then suggest some future work. Throughout this paper, we adopt $\rm H_{0}$ = 71 km s$^{-1}$ Mpc$^{-1}$, $\Omega_{\rm m}$ = 0.27 and $\Omega_{\rm DE}$ = 0.73. 

\section{VLA Observation \& Data Reduction} \label{sec:Observations}
\subsection{VLA Observations}
Very Large Array (VLA) observations at 1.4 GHz were retrieved from the National Radio Astronomy Observatory (NRAO) data archive. The data were all taken in wide field "pseudo-continuum mode" which consists of 25 MHz bandwidth observations acquired at two intermediate frequencies (IFs) centred at 1364.9 MHz and 1435.1 MHz. Each IF consists of seven 3.125 MHz channels and provides both left and right circular polarisation. Flux and phase calibration were tracked using flux calibrators (0137+331 or 3C 48) and a phase calibrator (0503+020), respectively. The summary of each archival VLA 1.4 GHz radio continuum observation is provided as follows:

\begin{enumerate}
\item BnA-array\footnote{Project ID: AN0109, PI: K. Nakanishi} data (hereafter, B-array) were obtained in June 9$^{\rm th}$-10$^{\rm th}$ 2002, that have a sensitivity as reported in the observing log of $\sim$ 0.052 mJy beam$^{-1}$ with on source target durations worth a total observing time of 7.8 hours, where data were recorded every 10 s. 
\item A-array\footnote{Project ID: AB1199, PI: A. Berciano Alba} data were obtained on February 5$^{\rm th}$-6$^{\rm th}$ and 10$^{\rm th}$-11$^{\rm th}$ 2006, that have a sensitivity as reported in the observing log of $\sim$ 0.04 mJy beam$^{-1}$ with on source target durations worth a total observing time of 12.8 hours, where data were recorded every 3.3 s.
\end{enumerate}

\subsection{VLA Data Reduction} \label{subsec:VLA Data Reduction}
The radio continuum data reduction and analysis were entirely carried out using the NRAO Astronomical Image Processing System (\AIPS\/) package \citep{2003ASSL..285..109G}. Our data reduction followed the standard calibration procedures that include data inspection, exploring visibilities, flagging, phase and flux calibrations. For both B-array and A-array, removal of corrupted data and calibration were performed in a similar manner. The final calibration solutions were applied individually to the different epochs.  We combined the two fully calibrated UV data sets and used wide field imaging techniques to image a large field of view beyond the primary beam.  We put facets on all bright sources within one square degrees of the pointing centre to properly account for flux and diminish sidelobe contamination.
 
Unlike the reduction procedures of \cite{2010A&A...509A..54B} of this cluster for their studies of sub-mm emissions from galaxies, we treated the B-array observations as single epoch. No self-calibration was applied. The noise in our combined image was found to be comparable to previously reduced data in the literature. Furthermore, no convolutional artifacts were found around the point sources. The A-array observations were run during the upgrade of the VLA with a mix of EVLA with VLA antennas used during the observing period. As a result, we discarded data from all three EVLA antennas because they could not be calibrated with the VLA antennas.

The A-array data set has higher resolution and is thus quite sensitive to point sources and small structures. Additionally, it dominates the total integrated time on source. As a result, any mapping parameters for the final radio map are dictated by the A-array. Facets for wide field imaging were based on bright sources found by \AIPS\ SETFC routine, resulting in 61 facets that were auto-generated for the targeted field coverage. Each facet was set to the same pixel scaling of 0.3" and 4096$\times$4096 pixels in size.  For all fields, during the cleaning process, cleaned regions were limited to each bright source by putting a clean box around it. The final clean maps have a convolved beam size (HPBW) of 1.99"$\times$1.63" at a position angle of 0.56. The final flattened radio map was imaged within a field of view coverage of 0.75$^{\circ}$$\times$0.75$^{\circ}$ with a pixel scaling of 0.3" and has an RMS noise level of 11.98 $\upmu$Jy beam$^{-1}$ as measured using \AIPS\/ TVSTAT. We define our detection limit for this image as being 36 $\upmu$Jy, which is 3 times the RMS found in a single beam. 

\subsection{Radio Sources Extraction and Cataloguing}
Our radio source catalogue was generated using the \AIPS\ task SAD by running this routine in the signal-to-noise mode (S/N$\geq$3). The \AIPS\ task RMSD was used to create an RMS noise map where calculation of the noise map was performed within a 30" diameter circular aperture corresponding to 100 neighbour pixels \cite[see e.g.,][]{2010ApJS..188..178M,2012ApJS..202....2W}. Source classification and flux were assigned based on the method adopted by \cite{2008AJ....136.1889O} using the best-fit major axis value. For the resolved sources (i.e., sources that have lower-limit best-fit major axis greater than zero), the total flux densities are directly assigned to the sources while peak flux densities are equal to the total flux densities for the unresolved sources. \\

We re-measured the 1.4 GHz flux density of fifteen sources published in \cite{2010MNRAS.401.2299W}  to verify the quality of our measurements. Within the estimated uncertainty range, we have found a very good consistency between the two measurements with a mean offset of 1.6 $\upmu$Jy and a dispersion of 5.5 $\upmu$Jy (i.e., within the RMS noise of our map).  

\section{Spitzer Observation \& Photometry} \label{Spitzer Observation and Photometry}
\subsection{Observations}
A Multiband Imaging Photometer for $Spitzer$ \citep[MIPS;][]{2004ApJS..154...25R}  24 $\upmu$m imaging mosaic for this field was retrieved from the $Spitzer$ Enhanced Imaging Products (SEIP\footnote{irsa.ipac.caltech.edu/data/SPITZER/Enhanced/Imaging/}) data archive. This high level product\footnote{ADS/IRSA.Atlas$\#$2013/0325/072424$\_$12320} consists of a combination of multiple programs and has a field of view  of $\sim$ 0.3$^{\circ}$$\times$0.3$^{\circ}$.  The mean integration time of the MIPS 24 $\upmu$m super mosaic imaging is 1637 seconds per pixel. The MIPS 24 $\upmu$m data has a median pixel scaling of 2.45"  and a mean FWHM of 5.9". The  RMS noise of the image was $\sim$ 26 $\upmu$Jy.

Additionally, $Spitzer$ super mosaics of the Infrared Array Camera \citep[IRAC;][]{2004ApJS..154...10F} at 3.6 $\upmu$m, 4.5 $\upmu$m, 5.8 $\upmu$m, 8.0 $\upmu$m were also retrieved. IRAC observations were used for sample selection and matching.  The IRAC images have a median pixel scaling of $\sim$ 1.2" and the image mean FWHM is 1.66", 1.72", 1.88", and 1.98", respectively. The  RMS noise of the images were $\sim$ 10 $\upmu$Jy.

\subsection{Far-Infrared Data at Longer Wavelength} 
No $Spitzer$ MIPS super mosaics imaging at 70, 160 $\upmu$m are available in the SEIP data archive for the cluster. However, $Spitzer$ MIPS imaging at 24, 70 and 160 $\upmu$m for the cluster are available in the Spitzer Heritage Archive (SHA\footnote{sha.ipac.caltech.edu/applications/Spitzer/SHA/}) but they have shallower depth and smaller field than the 24 $\upmu$m images. Due to these limitations, none of the confirmed cluster sources were matched with our 24 $\upmu$m catalogue. We have examined data from the recently available {\it Herschel} observations in the field for this cluster which consists of both PACS/PACS Evolutionary Probe (PEP) \citep{2011A&A...532A..90L, 2013A&A...553A.132M} and the SPIRE/{\it Herschel} Multi-tiered Extragalactic Survey (HerMES) \citep{2012MNRAS.424.1614O,2012MNRAS.419..377S}. The PACS data do not fully cover our field of view and we were only able to detect two sources in the SPIRE observations. As the {\it Herschel} data are not expected to significantly modify our results and due to the small number of sources with data, we did not attempt any further analysis using this data. The implications for only including 24 $\upmu$m observations are further discussed in \S \ref{Caveats}.

\subsection{Photometry}  
Optical coordinates (see \S \ref{sec:Sample Selection}) were used to perform aperture photometry on the $Spitzer$ observations. Aperture photometry was done using APEX \citep[][$Spitzer$ MOPEX]{2005PASP..117.1113M}.  The aperture size was set to a radius of 5.31"  in the MIPS 24 $\upmu$m imaging, which is the optimal aperture size suggested for point sources based on the pixel sampling of the MIPS 24 $\upmu$m PSF\footnote{irsa.ipac.caltech.edu/data/SPITZER/docs/}. Clusters sources  are expected to be essentially unresolved at this redshift in the MIPS 24 $\upmu$m data.  We re-measured the MIPS 24 $\upmu$m flux density of the sources published in \cite{2010MNRAS.401.2299W}  to verify the quality of our measurements.  For 15 sources published in \cite{2010MNRAS.401.2299W},  we have found a very good consistency between the two measurements with a mean offset of 16.8 $\upmu$Jy and a dispersion of 46 $\upmu$Jy.  

\section{Sample Selection} \label{sec:Sample Selection}
Our galaxy sample was drawn based on  spectroscopic data of the cluster MS0451-03 \citep{2007ApJ...671.1503M,2011ApJ...741...98C} and includes 350 cluster sources.  Cluster membership was determined through a ``shifting-gapper" analysis \cite[see,][]{2014ApJ...786...30C} similar to \cite{1996ApJ...473..670F} which determined membership through analysis of the radius-velocity diagram. Apart from the confirmed cluster galaxies, our sample also contains 1107 field galaxies that have secure spectroscopic redshift. 

Photometry was performed on the MIPS 24 $\upmu$m image for all spectroscopic sources in our sample appearing in those images.  Sources outside of the MIPS  24 $\upmu$m image were removed from our sample along with sources with flux densities  detected at $<$ 3$\upsigma$ level. The final catalogue of sources with secure spectroscopic redshifts and IR photometry comprises 155 cluster galaxies and 479 field galaxies. 
 
We matched the radio sources to this catalogue. A matching radius of 2" was used.  Any unambiguous identifications were discarded from the final catalogue. We were able to securely match 12 out of 155 cluster member galaxies and 27 out of 479 field galaxies.

Our full sample of matched sources is presented in Table \ref{tab:flux-lum-cluster} for cluster galaxies and Table \ref{tab:flux-lum-field} for the field galaxies. The tables include ID, RA, DEC, redshift, radius, 1.4 GHz flux density and its uncertainty, radio luminosity, MIPS 24 $\upmu$m flux density and its uncertainty, and IR luminosities.  As in \cite{2001ApJ...554..803Y} and \cite{2004ApJ...600..695R}, we have not a priori excluded AGN sources so our selection is a heterogeneous sample of star forming galaxies and AGN. The derivation of the intrinsic properties is presented in \S \ \ref{sec:Method}. 

\section{Methods} \label{sec:Method}
\subsection{The Radio Luminosity}
We converted the integrated radio flux densities ($\rm S_{1.4\rm GHz}$) into rest frame radio luminosities ($ \rm L_{1.4\rm GHz}$) using:  

\begin{equation}
\rm L_{1.4GHz}(W \ Hz^{-1}) = \left(\frac{4\pi [D_{L}(z)]^{2}}{(1+z)^{1-\alpha}}\right) \times S_{1.4\rm GHz}
\label{eq:L1.4}
\end{equation}
where S$_{1.4\rm GHz}$ is the flux density at 1.4 GHz in Jy, and D$_{L}(z)$ is the luminosity distance at the redshift of the sources. The $K$-correction 1/(1+z)$^{(1-\alpha)}$ consists of 1/(1 + z)$^{-\alpha}$ and 1/(1 + z) terms that are the ``colour" and bandwidth correction, respectively \citep{2003ApJS..146..267M}. The radio spectral index $\alpha$ is the power law slope of the synchrotron radiation, and is defined as S$_{\upnu}$$\sim$$\upnu ^{- \alpha}$. We assumed $\alpha$ $\sim$ 0.8 for normal star-forming  galaxies of \cite{1992ARA&A..30..575C}. 

\subsection{The Infrared Luminosities} \label{The Infrared Luminosities} 
Since we do not have sufficient information to fit  templates to individual galaxies, we used a simple recipe derived from \cite{2009ApJ...692..556R} for converting between observed 24 $\upmu$m flux density and  L$_{24 \rm\upmu m}$ based on the best-fit SFR calibration.  This is based on Equation 10-14 and Table 1 from  \cite{2009ApJ...692..556R}.  The conversion is as follows:

\begin{multline}
L{(24 \rm\upmu m, L_\odot)} = \frac{\rm 10^{A(z) + B(z)  \times (log(4 \pi D_{L}^{2} \ f_{24, obs}) - 53)}}{ 7.8 \times 10^{-10}}; \\ {\rm for,} \ 6\times10^{8} L_{\odot} \leq L_{24 \rm\upmu m} \leq 1.3\times10^{10} L_{\odot}
\label{eq:L24net01}
\end{multline}

\begin{multline}
L{(24 \rm\upmu m, L_\odot)} = \left(\frac{\rm 10^{A(z) + B(z)  \times (log(4 \pi D_{L}^{2} \ f_{24, obs}) - 53)}}{ 7.8 \times 10^{-10} \times (7.76 \times 10^{-11})^{0.048}}\right)^{0.954}; \\ {\rm if,} \ L_{24 \rm\upmu m} > 1.3\times10^{10} L_{\odot}
\label{eq:L24net02}
\end{multline}

where $D_{L}$ is the luminosity distance in cm and $f_{\rm 24, obs}$ is the flux density at 24 $\upmu$m in Jy. The coefficients A(z) and B(z) are redshift-dependent and can be obtained by interpolating the values in Table 1 of  \cite{2009ApJ...692..556R}; or see values are provided in our Table A1 in the Appendix \ref{app:equations}. 

In addition to L$_{24 \rm\upmu m}$, we also calculate the IR luminosities L$_{\rm TIR}$, L$_{60 \rm\upmu m}$, and L$_{\rm FIR}$ to allow a comparison of our measurements to other works. We adopt the definition from \cite{2009ApJ...692..556R} that L$_{\rm TIR}$ is the luminosity between L(TIR; 8--1000 $\upmu$m) and we adopt the definition of L$_{\rm FIR}$ from \cite{1985ApJ...298L...7H} as L(FIR; 42--122 $\upmu$m).  We provide the details of these transformations in Appendix \ref{app:equations}.

We also attempted to determine L$_{\rm TIR}$ based on fitting the spectral energy distribution (SED) using CIGALE\footnote{http://cigale.oamp.fr/} \citep{2009A&A...507.1793N}. The photometric catalogues were fit to synthetic template spectra compiled from \cite{2005MNRAS.362..799M} that incorporate the dust emission model of \cite{2002ApJ...576..159D}. The total infrared luminosities were then derived from the parameters of the best-fit model. Due to lacking suitable measurements in bandpasses that are redward of 24 $\upmu$m, the scarcity of data points, and the large parameter space of models, our measurements of the total IR luminosity L$_{\rm TIR}$ were poorly constrained. We thus limited any further analysis to the empirically determined L$_{\rm TIR}$.  

\subsection{The IR to Radio Luminosity Ratio} 
We characterised the quantitative measure of the far-infrared-radio relation by calculating the median logarithmic ratio of IR and radio luminosity ($q$). The luminosity ratio was estimated using the commonly used equation of \cite{1985ApJ...298L...7H} as follows:  

\begin{equation}
 q =\rm log \left(\frac{L}{3.75 \times 10^{12} \ \rm W  }\right) -  log \left(\frac{L_{\rm 1.4GHz}}{\rm W  \ Hz^{-1}}\right)
\end{equation}  
where L$_{\rm 1.4GHz}$ is the rest frame radio luminosity calculated from Equation \ref{eq:L1.4} in W Hz$^{-1}$. L is our infrared luminosity in W. The subscript of $q$ indicates which infrared luminosity is being used (i.e., q$_{\rm TIR}$ is for L$_{\rm TIR}$).  For calculating $q_{24}$, the constant used to normalise the infrared luminosity was $1.25 \times 10^{13}$.

\section{Results} \label{sec:Results}
Our primary aim is to compare the infrared-radio relationship in an intermediate redshift cluster to nearby clusters, and to do so, we will be comparing our results to the lower redshift measurements of  \cite{2004ApJ...600..695R}.  As their results are reported in L$_{60 \rm\upmu m}$ and q$_{\rm FIR}$ within L(FIR; 42--122 $\upmu$m), we transform our L$_{24 \rm\upmu m}$ results into comparable bands. We report the luminosities for all the cluster sources in Table \ref{tab:flux-lum-cluster}  and field sources in Table \ref{tab:flux-lum-field}.

\subsection{Far-Infrared-Radio Relation}
\begin{figure} 
\begin{center} 
\includegraphics[scale=0.36]{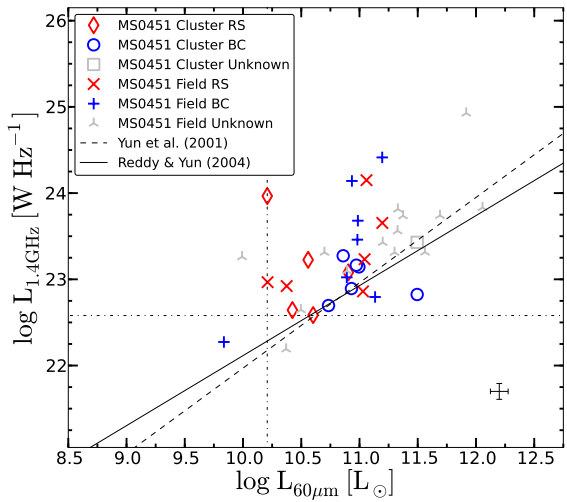}
\caption{The 20 cm radio continuum luminosity (L$_{\rm 1.4GHz}$) against the IR luminosity (L$_{\rm 60 \upmu m}$). Red colour indicates red sequence (RS) galaxies while blue colour represents blue cloud (BC) galaxies.  RS field galaxies are plotted in cross symbols, while BC field galaxies are drawn in plus symbols. Grey colour represents sources with unknown photometric classification. The solid and dashed lines indicate the formal linear least-square fit (log L$_{\rm 60 \upmu m}$ = 8.92 luminosity cutoff) of the low redshift cluster galaxies from \protect\cite{2004ApJ...600..695R} and field galaxies relation in Equation 4 of \protect\cite{2001ApJ...554..803Y}. The error bars correspond to average 1$\upsigma$ errors.}  
\label{fig:ir-radio-emp} 
\end{center} 
\end{figure}

The relationship between the rest frame radio luminosity at 1.4 GHz (L$_{1.4\rm GHz}$) against the IR luminosity (L$_{60 \rm\upmu m}$) is shown in Figure \ref{fig:ir-radio-emp}. The solid line indicates the formal linear least-square fit of the cluster galaxies of \cite{2004ApJ...600..695R} while the field relation from \cite{2001ApJ...554..803Y} is drawn using the dashed line. Most of our cluster sources are consistent with these relationships. As indicated in the dash-dotted lines in Figure \ref{fig:ir-radio-emp}, our cluster galaxies IR luminosity and radio luminosity lower limits are log L$_{60 \rm\upmu m}$ = 10.21 and log L$_{\rm 1.4GHz}$ = 22.6, respectively. For comparison, our lower limits are higher than the low redshift cluster galaxies (log L$_{60 \rm\upmu m}$ = 8.92, log L$_{\rm 1.4GHz}$ = 20.47) of \cite{2004ApJ...600..695R}. 

We split our sample into blue cloud (BC) and red sequence (RS) galaxies following the definition in \cite{2011ApJ...741...98C} for both cluster and field samples. In all Figures, red and blue colours indicate sources that have secure photometric measurements from our imaging data, while grey colour represents sources with unknown photometric classification. The number of sources that have secure photometric classification is 11 of 12 cluster sources and 13 of 27 field sources. For any values given for these populations, we only use sources with secure photometric measurements. For all other values reported in this paper, we calculate them based on the full cluster or field sample.

\begin{figure} 
\begin{center} 
\includegraphics[scale=0.36]{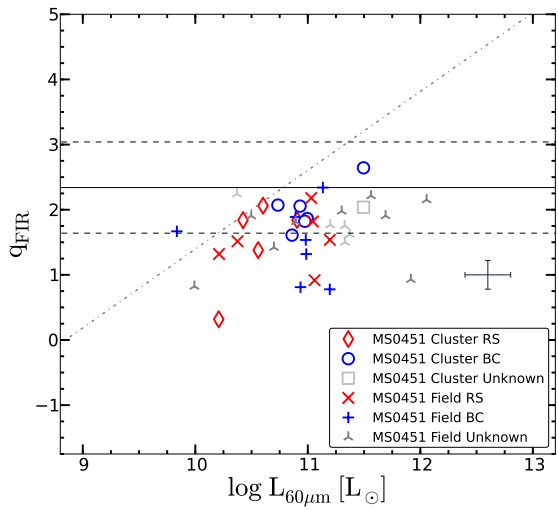}
\caption{The logarithm of the far-IR luminosity to 1.4 GHz radio continuum luminosity ratio ($q_{\rm FIR}$) versus the IR luminosity (L$_{60 \rm\upmu m}$). The nominal value of $q_{\rm FIR}$ for field galaxies ($q_{\rm FIR}$=2.34) is plotted in the solid black horizontal line. The criteria for both delineating the radio-excess ($q_{\rm FIR}$$\leq$1.64) and IR-excess ($q_{\rm FIR}$$\geq$3.04) are shown in the dashed lines. The error bars correspond to average 1$\upsigma$ errors. The solid gray line represents our sample limiting flux.} 
\label{fig:ir-q-emp} 
\end{center} 
\end{figure}

In Figure \ref{fig:ir-q-emp}, we present the far-IR luminosity (L$\rm _{FIR}$) to radio luminosity ($ \rm L_{1.4\rm GHz}$) ratio ($q_{\rm FIR}$) versus  L$_{60 \rm\upmu m}$. The solid gray line delineates our sample limiting magnitude. The mean $q_{\rm FIR}$ for the cluster population is $q_{\rm FIR}$ = 1.80$\pm$0.15 with a dispersion of 0.53, while field galaxies have $q_{\rm FIR}$ =  1.62$\pm$0.09 with a dispersion of 0.45.  Our cluster value is consistent with \cite{2004ApJ...600..695R} value of $q_{\rm FIR}$ = 2.07$\pm$0.07 with a dispersion of 0.74. Our field value is lower, but due to the large intrinsic dispersion, we cannot firmly comment on its inconsistency with values found in similar works such as \cite{2001ApJ...554..803Y} of $q_{\rm FIR}$ = 2.34$\pm$0.01 with a dispersion of 0.26. In addition to \cite{2004ApJ...600..695R} and \cite{2001ApJ...554..803Y}, our mean $q_{\rm FIR}$ values are comparable with other works \citep{1995AJ....109.1582A,2001AJ....121.1903M,2009ApJ...694.1435M,2002A&A...384L..19G,2006ApJ...650..592K,2008ApJ...683..659S}; see Table \ref{tab:q-define} for values from each of the surveys.

We also compare our other measurements to other works as well.  For comparison, we found mean values of $q_{24}$ = 0.69$\pm$0.16 with a dispersion of 0.55 and $q_{\rm TIR}$ = 2.10$\pm$0.15 with a dispersion of 0.53 for cluster sources while $q_{24}$ = 0.52$\pm$0.09 with a dispersion of 0.46 and $q_{\rm TIR}$ = 1.92$\pm$0.09 with a dispersion of 0.45 for our field sources. A comparison of $q_{24}$ and $q_{\rm TIR}$ values in the field indicates that our values are consistent with previously published values of $q_{24}$ \citep{2004ApJS..154..147A, 2006ApJ...638..157M, 2008MNRAS.385.1143B,2008MNRAS.386..953I, 2009MNRAS.397.1101G} and $q_{\rm TIR}$ \citep{2009ApJ...698.1380M,2010MNRAS.402..245I,2010MNRAS.409...92J,2010A&A...518L..31I}.  

As for the low redshift environment, the field and cluster $q_{\rm FIR}$ are consistent, although with the large dispersions measured for our small sample, it is difficult to notice any significant differences.   The field sample shows a strong bias with redshift (with more excess objects at higher redshift) and as such, we exclude it from further analysis. The  q-value will be further discussed in \S \ref{sec:Discussions}.

In addition, Figure \ref{fig:ir-q-emp} shows that most sources lie within the adopted range defined as radio-excess ($q_{\rm FIR}$$\leq$1.64) and IR-excess ($q_{\rm FIR}$$\geq$3.04) by \cite{2001ApJ...554..803Y} as shown in the dashed lines.  A radio (IR) excess galaxy is defined to have at least five times greater radio (IR) flux than what is expected from the field galaxy at that redshift for a given far-IR luminosity.  We find that the percentage\footnote{Errors for percentages were calculated following \cite{1986ApJ...303..336G}} of cluster galaxies that have radio-excess are $27^{+23}_{-13}\%$. The number of radio-excess cluster members is higher than the  percentage found by \cite{2004ApJ...600..695R} ($11^{+1.7}_{-4.7}\%$) for low redshift clusters.

Out of the cluster population, 2 of 5 (40\%) RS galaxies are radio-excess sources.  In their sample, \cite{2004ApJ...600..695R} find 28\% of cluster early type galaxies show a radio-excess and  that all also display evidence of AGN activity.  For the BC cluster galaxies, we find 1 of 6 (16\%) radio-excess sources, which is consistent with the results of \cite{2001AJ....121.1903M} and \cite{2004ApJ...600..695R}. For these blue population, where $q_{\rm FIR}$ = 2.01$\pm$0.14 with a dispersion of 0.35, which is consistent with the field at low redshift and intermediate redshift.

\begin{figure} 
\begin{center} 
\includegraphics[scale=0.36]{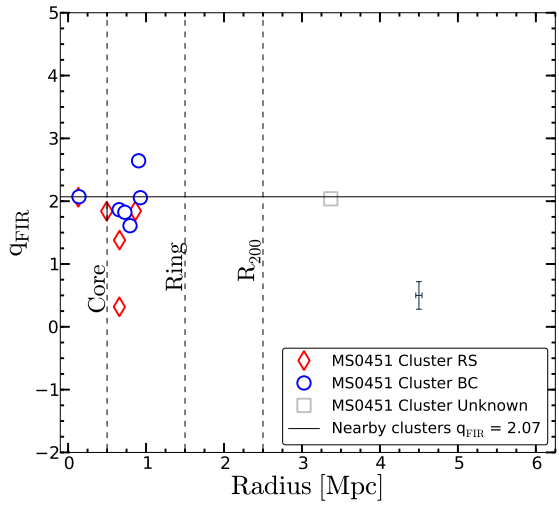}
\caption{Plot of the galaxy projected radius against the far-IR to radio luminosity ratio $q_{\rm FIR}$. The mean value of $q_{\rm FIR}$ for low redshift cluster galaxies of \protect\cite{2004ApJ...600..695R} is shown in the solid horizontal line ($q_{\rm FIR}$=2.07). Similar to \protect\cite{2004ApJ...600..695R}, the vertical dashed lines define the R$_{\rm Core}$, R$_{\rm Ring}$ as the projected cluster-centric distance at R $\leq$ 0.5 and between [0.5, 1.5] Mpc, respectively. The error bars correspond to average 1$\upsigma$ errors.} 
\label{fig:q-crad-emp} 
\end{center} 
\end{figure}

In Figure \ref{fig:q-crad-emp}, we plotted the far-IR and radio luminosity ratio ($q_{\rm FIR}$) as a function of the galaxy projected radius R in Mpc. In aiming to use radius as a proxy for local density, similar to \cite{2004ApJ...600..695R}, we defined R$_{\rm Core}$, R$_{\rm Ring}$ as the projected cluster-centric distance at R $\leq$ 0.5 Mpc and between [0.5, 1.5] Mpc, respectively.  \cite{2001AJ....121.1903M} and \cite{2004ApJ...600..695R} find a higher fraction of core galaxies that have a radio-excess, but we see no evidence of this in our sample drawn from one cluster.

\begin{landscape}
\begin{table}
\begin{center}
\caption{Properties of cluster galaxies that include the total IR luminosity (L$_{\rm TIR}$)  and rest frame radio luminosity (L$_{\rm 1.4GHz}$).} 
\begin{tabular}{l c c c c c c c c c c c c c} \hline 
 ID & RA          & DEC       & Redshift & $R$   & S$_{\rm 1.4}$ &  S$_{\rm 1.4}$-err  & log L$_{\rm 1.4}$   & S$_{\rm 24}$ & S$_{\rm 24}$-err  & log(L$_{\rm 24\upmu m}$)   & log(L$_{\rm 60\upmu m}$) & log(L$_{\rm FIR}$)   & log(L$_{\rm TIR}$)  \\ 
     &  (Degree) &  (Degree) &      &   (Mpc) & ($\upmu$Jy)            & ($\upmu$Jy)         & (W Hz$^{-1}$)          & (mJy) & (mJy)       &  (L$_{\odot}$) &   (L$_{\odot}$) &  (L$_{\odot}$) &   (L$_{\odot}$)      \\  \hline

1081 & 73.514153 & -2.988997 & 0.531  & 0.927 & 75.906 & 24.331 & 22.893 &	0.323 & 0.027 &		10.364 & 10.931 & 10.938 & 11.239 \\
1093 & 73.521263 & -2.994204 & 0.527  & 0.728 & 144.311 & 21.045 & 23.164 &	0.357 & 0.027 &		10.405 & 10.973 & 10.977 & 11.278 \\
1118 & 73.518738 & -3.003781 & 0.532  & 0.656 & 134.7 & 25.47 & 23.144 &	0.363 & 0.027 &		10.427 & 10.995 & 10.998 & 11.299 \\
1143 & 73.516792 & -2.990292 & 0.532  & 0.863 & 116.476 & 24.413 & 23.080 &	0.307 & 0.027 &		10.339 & 10.904 & 10.914 & 11.215 \\
1158 & 73.520401 & -3.000317 & 0.542 & 0.660 & 156.083 & 24.667 & 23.226 &	0.156 & 0.027 &		10.002 & 10.559 & 10.596 & 10.897 \\
1178 & 73.523178 & -2.988423 & 0.530 & 0.794 & 183.027 & 54.111 & 23.273 &	0.286 & 0.027 &		10.295 & 10.860 & 10.873 & 11.174 \\
1489 & 73.534897 & -3.054384 & 0.548  & 0.905 & 60.155 & 20.114 & 22.824 &	0.846 & 0.027 &		10.914 & 11.495 & 11.457 & 11.758 \\
1726 & 73.549652 & -3.044432 & 0.541  & 0.658 & 862.685 & 98.963 & 23.966 &	0.084 & 0.027 &		9.661 & 10.208 & 10.274 & 10.575 \\
1811 & 73.550758 & -3.016844 & 0.539  & 0.132 & 36.000 & 12.000 & 22.585 &	0.170 & 0.023 &		10.045 & 10.602 & 10.636 & 10.937 \\

2143 & 73.558327 & -3.033125 & 0.544  & 0.495 & 40.246 & 14.39 & 22.643 &	0.121 & 0.026 &		9.871 & 10.424 & 10.472 & 10.774 \\

7240 & 73.551170 & -3.016482 & 0.539  & 0.140 & 46.558 & 13.505 & 22.696 &	0.216 & 0.023 &		10.172 & 10.734 & 10.757 & 11.058 \\

8110 & 73.396484 & -3.020605 & 0.536  & 3.370 & 255.464 & 73.437 & 23.429 &	0.897 & 0.021 &		10.914 & 11.495 & 11.457 & 11.758 \\

\hline
\end{tabular} 
\label{tab:flux-lum-cluster} 
\end{center}
\end{table}
\end{landscape}

\begin{landscape}
\begin{table}
\begin{center}
\caption{Properties of field galaxies that include the total IR luminosity (L$_{\rm TIR}$)  and rest frame radio luminosity (L$_{\rm 1.4GHz}$).} 
\begin{tabular}{l c c c c c c c c  c c c c c} \hline 
 ID & RA          & DEC       & Redshift &  $R$   & S$_{\rm 1.4}$ &  S$_{\rm 1.4}$-err  & log L$_{\rm 1.4}$   & S$_{\rm 24}$ & S$_{\rm 24}$-err  & log(L$_{\rm 24\upmu m}$)   & log(L$_{\rm 60\upmu m}$) & log(L$_{\rm FIR}$)   & log(L$_{\rm TIR}$)  \\ 
     &  (Degree) &  (Degree) &       &     (Mpc) & ($\upmu$Jy)            & ($\upmu$Jy)         & (W Hz$^{-1}$)          & (mJy) & (mJy)       &  (L$_{\odot}$) &   (L$_{\odot}$) &  (L$_{\odot}$) &   (L$_{\odot}$)      \\  \hline
1046 & 73.513374 & -2.989720 & 0.578 &  0.930 & 1099.165 & 125.937 & 24.141 &	0.260 & 0.027 &		10.368 & 10.935 & 10.942 & 11.243 \\
1743 & 73.547653 & -3.042384 & 0.586 &  0.606 & 64.228 & 16.439 & 22.922 &	0.091 & 0.027 &		9.824 & 10.375 & 10.427 & 10.728 \\
1937 & 73.550400 & -3.057221 & 0.655 &  0.947 & 1530.335 & 260.189 & 24.414 &	0.314 & 0.027 &		10.621 & 11.195 & 11.181 & 11.482 \\
2472 & 73.570877 & -3.027037 & 0.617 &  0.639 & 307.426 & 28.24 & 23.654 &	0.359 & 0.026 &		10.622 & 11.196 & 11.182 & 11.483 \\
2818 & 73.587166 & -2.987284 & 0.239 &  1.153 & 114.59 & 31.547 & 22.273 &	0.317 & 0.029 &		9.299 & 9.836 & 9.931 & 10.232 \\
2912 & 73.586548 & -2.998683 & 0.725 &  1.018 & 79.373 & 25.822 & 23.233 &	0.193 & 0.030 &		10.475 & 11.044 & 11.043 & 11.344 \\
3017 & 73.591164 & -2.990986 & 0.489 &  1.187 & 109.072 & 34.843 & 22.967 &	0.110 & 0.036 &		9.664 & 10.211 & 10.276 & 10.577 \\
3226 & 73.603012 & -2.980324 & 0.725 &  1.541 & 656.767 & 39.495 & 24.150 &	0.198 & 0.040 &		10.491 & 11.060 & 11.058 & 11.359 \\
3462 & 73.610458 & -2.947674 & 0.728 &  2.141 & 220.184 & 38.439 & 23.680 &	0.172 & 0.039 &		10.418 & 10.986 & 10.989 & 11.290 \\
3806 & 73.626404 & -2.999176 & 0.332 &  1.883 & 210.369 & 38.786 & 22.860 &	1.336 & 0.034 &		10.461 & 11.030 & 11.030 & 11.331 \\
3957 & 73.619141 & -2.937129 & 0.323 &  2.450 & 192.012 & 58.31 & 22.795 &	1.741 & 0.032 &		10.562 & 11.134 & 11.125 & 11.426 \\
5823 & 73.624580 & -2.969661 & 0.447 &  2.084 & 153.757 & 41.791 & 23.023 &	0.472 & 0.033 &		10.326 & 10.891 & 10.902 & 11.203 \\
6380 & 73.580307 & -2.921304 & 0.619 &  2.286 & 245.952 & 45.472 & 23.561 &	0.453 & 0.013 &		10.750 & 11.327 & 11.302 & 11.603 \\
6638 & 73.648750 & -2.975052 & 0.447 &  2.526 & 390.0 & 50.734 & 23.427 &	0.834 & 0.021 &		10.625 & 11.199 & 11.185 & 11.486 \\
7372 & 73.573524 & -2.971213 & 0.777 &  1.199 & 113.395 & 31.084 & 23.460 &	0.149 & 0.027 &		10.415 & 10.982 & 10.986 & 11.287 \\
8060 & 73.394081 & -3.050711 & 0.491 &  3.513 & 213.045 & 73.092 & 23.261 &	0.074 & 0.022 &		9.450 & 9.992 & 10.074 & 10.375 \\
8134 & 73.402657 & -3.035359 & 0.715 &  3.258 & 312.974 & 74.698 & 23.814 &	0.329 & 0.023 &		10.753 & 11.330 & 11.306 & 11.607 \\
8415 & 73.586739 & -3.150578 & 0.505 &  3.199 & 224.847 & 63.553 & 23.313 &	0.240 & 0.013 &		10.138 & 10.699 & 10.725 & 11.026 \\
8575 & 73.623688 & -2.891933 & 0.900 &  3.326 & 2337.128 & 397.688 & 24.927 &	0.454 & 0.022 &		11.324 & 11.916 & 11.845 & 12.146 \\
8578 & 73.655594 & -2.971230 & 0.802 &  2.702 & 247.537 & 53.255 & 23.832 &	0.864 & 0.021 &		11.459 & 12.056 & 11.973 & 12.274 \\
8581 & 73.659332 & -2.969704 & 0.219 &  2.794 & 115.513 & 41.055 & 22.191 &	1.068 & 0.022 &		9.818 & 10.370 & 10.422 & 10.723 \\
8583 & 73.648529 & -2.962799 & 0.590 &  2.636 & 156.628 & 50.165 & 23.315 &	0.788 & 0.025 &		10.979 & 11.562 & 11.519 & 11.820 \\
8608 & 73.670662 & -2.991703 & 0.447 &  2.900 & 183.367 & 62.913 & 23.099 &	0.474 & 0.022 &		10.327 & 10.893 & 10.903 & 11.204 \\
8631 & 73.664940 & -3.052018 & 0.794 &  2.840 & 205.325 & 56.57 & 23.741 &	0.477 & 0.021 &		11.104 & 11.691 & 11.637 & 11.938 \\
8642 & 73.693115 & -3.053544 & 0.280 &  3.464 & 188.581 & 69.843 & 22.646 &	0.749 & 0.025 &		9.942 & 10.497 & 10.539 & 10.840 \\
8773 & 73.578003 & -2.830110 & 0.491 &  4.275 & 239.729 & 85.027 & 23.312 &	0.789 & 0.026 &		10.725 & 11.301 & 11.279 & 11.580 \\
8813 & 73.656097 & -2.904510 & 0.756 &  3.564 & 229.033 & 67.389 & 23.736 &	0.312 & 0.023 &		10.796 & 11.374 & 11.346 & 11.647 \\
\hline
\end{tabular} 
\label{tab:flux-lum-field} 
\end{center}
\end{table}
\end{landscape}

\subsection{Potential Caveats}  \label{Caveats}  
Before interpreting our findings, there are some caveats related to the derived luminosities which need to be considered.
 
\subsubsection{Far-Infrared-Radio Relation: Presence of AGN}

\begin{figure}
\begin{center} 
\includegraphics[scale=0.35]{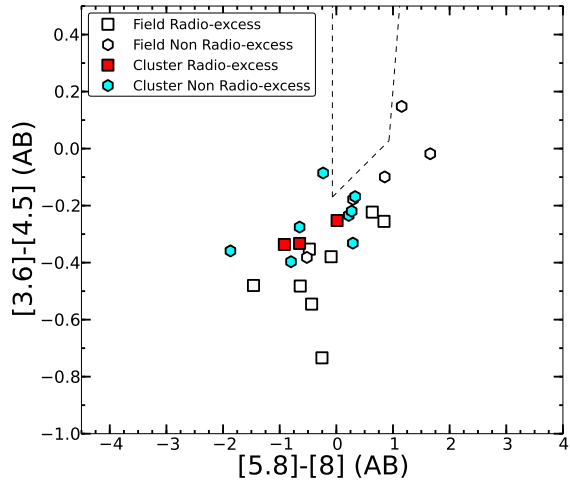}
\caption{Plot of AGN indices. IRAC colour-colour diagnostic for AGN. } 
\label{fig:irac_test2} 
\end{center} 
\end{figure}

For comparison purposes and conformity to the previous work in the literature \citep{2004ApJ...600..695R}, we do not a priori exclude AGN, but analyse both star-forming galaxies and AGN together.  It has been established that faint radio populations are mostly found to be composed of star-forming galaxies and radio-quiet AGN \cite[e.g.][]{2004NewAR..48.1173J}. It is also acknowledged that there is generally a contribution to the net radio flux from an AGN which can affect the observed relationship. However, we do note that the optical spectra for our sources do not show evidence of AGN activity based on the lack of broad-line sources. We then followed the method of \cite{2005ApJ...631..163S} to check for AGN contamination using IRAC colour-colour plots which is shown in Figure \ref{fig:irac_test2}. We particularly adopted the formulation in AB mag system by  \cite{2010ApJ...719..790M} as shown in their Figure 4. We find no clusters sources that show any indication of being an AGN according to the classification as defined by \cite{2005ApJ...631..163S,2010ApJ...719..790M}. The implications of AGN in our sample are further discussed in \S \ref{sec:Discussions}.

\subsubsection{IR Luminosity Derived From 24 $\mu$m}
Observations at 24 $\upmu$m may not  be providing an unbiased estimator of the star formation since the peak of the IR SED tracing the cold dust component peaks between 60 $\upmu$m and 170 $\upmu$m \citep{2003A&A...409..907P}.  Furthermore, 24 $\upmu$m data may be affected by dust heating from older stellar populations \citep{2010ApJ...714.1256C}, although the 24 $\upmu$m flux is going to be dominated by the warm dust component.

In addition, there is an order a magnitude correction required to convert  24 $\upmu$m flux to  total IR flux, which small scatter in the relationship can contribute large uncertainties. However, \cite{2001ApJ...549..215D} argued that 20-42 $\upmu$m is essentially a good tracer of the bulk of dust emission and hence can be a robust recent star formation indicator. Recent studies of the relationship in the field have also found consistent results between mid-infrared MIPS 24 $\upmu$m and MIPS 70 $\upmu$m results \cite[e.g.][]{2004ApJS..154..147A, 2008MNRAS.385.1143B}. In addition, \cite{2011ApJ...732..126M} find that 24 $\upmu$m observations are a sufficient tracer of the total IR luminosity of galaxies for galaxies with L$_{\rm 24\upmu m}$$<$10$^{12}$  L$_{\odot}$, which includes all of our cluster sample. 

\cite{2013MNRAS.431.1956G} combined {\it Spitzer} and {\it Herschel} data and noted that an inclusion of 24 $\upmu$m wavelength is essential in order to robustly derive the total IR luminosity for nearby star-forming galaxies. In addition, for luminous galaxies at z $<$ 1.3, measurement based on mid-IR is in agreement with those measured directly with {\it Herschel}, as already shown by  \cite{2011A&A...533A.119E}.

\subsubsection{Small Sample Size}
Given the fact that we have studied only one cluster containing a relatively small number of members, it is important to note that the current results may suffer from small sample size.  In our sample, we did not detect the brightest cluster galaxy (which is obscured by a foreground galaxy) while \cite{2004ApJ...600..695R} have 4 cDs in their sample.

\section{Discussions} \label{sec:Discussions}
Among the low redshift cluster studies, the far-infrared-radio relationship in rich cluster galaxies is characterised by a lower value of $q_{\rm FIR}$ as compared to the field that is indicative of an excess of radio emission.  Cluster environmental effects are believed to drive these observations \citep{1995AJ....109.1582A,1997MNRAS.290....1R}.  \cite{1995AJ....109.1582A} postulated that the ISM of galaxies in rich clusters is being compressed via ram pressure as the galaxies move through the ICM resulting in greater radio emission. More recent work has further examined various models to shed some light on the causes of enhanced  radio emission  that seems globally present in cluster galaxies \citep{2001AJ....121.1903M,2004ApJ...600..695R,2009ApJ...694.1435M}. The most common scenarios include thermal pressure compression by the ICM and the ram pressure stripping  of ISM, which are both likely to augment the galactic magnetic field. 

If we compare our results with the study of low redshift clusters by \citealp{2004ApJ...600..695R}, we find the following: (1) the intermediate redshift cluster sample has a lower value of $q_{\rm FIR}$ = 1.80$\pm$0.15 with a dispersion of 0.53, but one that is consistent within the dispersion with the low redshift clusters where $q_{\rm FIR}$ = 2.07$\pm$0.07 with a dispersion of 0.74; (2) the fraction of radio-excess objects in clusters at intermediate redshifts is greater than in the low redshift ($27^{+23}_{-13}\%$ to $11\%$); (3) We find no preference for radio excess objects in the cluster core.   However, we caution that any results from this work may suffer from the small sample size studied here. 

Our measurement of $q_{\rm TIR}$ in the field at intermediate redshift is consistent with previous work \citep{2009ApJ...698.1380M,2010MNRAS.402..245I, 2010MNRAS.409...92J,2010A&A...518L..31I}, and little or no evolution has been reported in the evolution of the $q_{\rm TIR}$ value in the field at these redshifts \cite[e.g.][]{2010ApJS..186..341S,2010ApJ...714L.190S,2010A&A...518L..31I,2011MNRAS.410.1155B}. However, the significant increase in the scatter in the $q_{\rm FIR}$ value is inconsistent with these works as well as the scattered measured by \citealp{2001ApJ...554..803Y} at low redshift. The high scatter that we are measuring may be due to preferentially selecting radio bright objects especially at higher redshift. For this reason, we will focus on the observed fraction of radio-excess objects seen in the cluster and the behaviour of blue galaxies in the cluster.

In the intermediate redshift cluster, we observe a higher fraction of radio-excess galaxies as compared to lower redshift. Two thirds of the sources classified as radio-excess sources are red sequence galaxies.  \cite{2004ApJ...600..695R} found that 9 of 13 excess sources at low redshift had AGN signatures and were early-type galaxies. From visual inspection of their spectra and analysis of their mid-infrared colours, our two red galaxies do not show broad emission lines or other telltale AGN signatures, but their large offset from the relationship may be indicative of nuclear activity.   As can be seen in Figure \ref{fig:ir-q-emp}, most of the blue galaxies appear normal and have a very small scatter in their $q_{\rm FIR}$ values. We find one blue galaxy showing a radio-excess. 

In Figure \ref{fig:q-crad-emp}, we do not see any strong radial trends against the $q_{\rm FIR}$ values. However, we notice that there is a significant scatter within the ring galaxies. Overall, inspection of the median $q_{\rm FIR}$ value indicates that the far-infrared-radio relation for cluster blue galaxies at $z$ $\sim$ 0.5 is similar to the cluster sample at low redshift.  

At low redshift clusters, one interpretation of the presence of radio-excess within the virial radius can be the results of interstellar medium (ISM) stripped off via ram pressure exerted by infalling galaxies.   We do not see any strong environmental effects in the value of $q_{\rm FIR}$ at these redshifts, although we do see a greater percentage of excess objects.  It is possible that the quenching mechanism is different than it is at low redshift although a larger sample would be required to verify this.  Follow-up observations of these excess objects found at intermediate redshift clusters would also help in exploring their properties. 

However, we note the radio spectral index $\alpha$ depends on frequency and galaxy properties and thus using a single spectral index may also alter the value of $q$ \cite[e.g.][]{2011MNRAS.410.1155B}. However, the majority of work of this kind has been using the standard spectral index for normal star-forming galaxies; $\alpha$ $\sim$ 0.8 of \cite{1992ARA&A..30..575C}. Recent results from the local infrared luminous galaxies of \cite{2013ApJ...768....2M} found that the mid-infrared and radio properties of star-forming galaxies, particularly for those compact starburst galaxies, tend to have a flatter ($\alpha$ $\sim$ 0.5) spectral index.   This may be a concern for galaxies in clusters as \cite{2006ApJ...636L..13C,2011ApJ...741...98C} find an increase in compact galaxies in clusters at these redshifts. The recently upgraded VLA correlator now makes it possible to measure the spectral index over a wide bandwidth within a single observation, so it will be a powerful tool for determining the IR-radio relation for large field and cluster samples in the future.

\section{Conclusion} \label{sec:Conclusion}
We have studied the far-infrared-radio relation in the massive galaxy cluster MS0451-03 and investigated, for the first time, how this relationship behaves at intermediate redshift between the field and a high-density cluster environment. We have constructed the far-infrared-radio relationship of star-forming galaxies using deep VLA and $Spitzer$ archival data. We have measured the rest frame radio luminosity at 1.4 GHz and the total IR luminosity ratios for both sample of confirmed MS0451-03 cluster and field galaxies. 

We find that the far-infrared-radio relationship for distant cluster populations $(q_{\rm FIR}$ = 1.80$\pm$0.15 with a dispersion of 0.53) is in agreement with those measured in  low redshift clusters ($q_{\rm FIR}$ = 2.07$\pm$0.07 with a dispersion of 0.74).  We do find an evidence for a cluster enhancement of the radio excess sources with the value in MS0451-03 ($27^{+23}_{-13}\%$) being significantly higher than that in low redshift clusters ($11^{+1.7}_{-4.7}\%$). In addition, the far-infrared-radio relationship for blue galaxies, where $q_{\rm FIR}$ = 2.01$\pm$0.14 with a dispersion of 0.35, is consistent with the predicted value from the field relationship, although these results are based on a sample from a single cluster. We find one radio-excess galaxy among the blue star forming galaxies and two RS galaxies with radio-excess.  Unlike low redshift galaxies, our galaxies do not show any evidence of AGN activity, but further observations will be needed to confirm the nature of these objects.   We do not find any trends with radius for radio-excess populations within this one cluster. 
 
In the future work, we will expand our sample of distant and rich clusters to further explore the relationship at intermediate redshift. The subsequent analysis of the extended sample shall provide higher statistics and will allow further confirmation of differences seen between low and intermediate redshift.  

\section*{Acknowledgements} 
We would like to thank the anonymous referee for his/her thorough review of our manuscript and the valuable comments. We thank Dr. Elodie Giovannoli for her helpful discussions and suggestions. SMR and MV were supported by the South African Square Kilometre Array Project (SKA SA) and the National Research Foundation (NRF). SMR wishes to thank the South African Astronomical Observatory NRF/SAAO for their support. KMH's research has been supported by the South African Research Chairs Initiative (SARChI) of the Department of Science and Technology (DST), the SKA SA and the NRF. 

We are grateful to the W. M. Keck Observatory for the observations. The National Radio Astronomy Observatory is a facility of the National Science Foundation operated under cooperative agreement by Associated Universities, Inc. This research has made use of the NASA/ IPAC Infrared Science Archive, which is operated by the Jet Propulsion Laboratory, California Institute of Technology, under contract with the National Aeronautics and Space Administration: ADS/IRSA.Atlas$\#$2013/0325/072424$\_$12320. 

\bibliographystyle{mn2e}
\bibliography{fir-radio}

\appendix
\section{Equations Used To Estimate IR Luminosities}  
\label{app:equations}

Additional materials are provided in this Appendix which consists of empirical relations used to estimate IR luminosities and a table that contains various $q$-values in the literature (See Table \ref{tab:q-define}).  

\subsection{IR Luminosities Inferred From Empirical Relations}

\subsubsection{Estimating total IR luminosity (L$_{\rm TIR}$)} 
Measurement of IR luminosity can be made through an empirical relation via either a single IR band or a combination of many IR bands. The total IR luminosity for our sources was solely determined through the MIPS 24 $\upmu$m empirical relation. 

\cite{2009ApJ...692..556R} compute total IR luminosity (L$_{\rm TIR}$) as described in \cite{2003AJ....126.1607S,1996ARA&A..34..749S}. The luminosity estimator of \cite{1996ARA&A..34..749S} is defined at $\lambda =8\mbox{--}1000\,\mu\mbox{m}$; L(TIR; 8--1000 $\upmu$m) and L$_{\rm TIR}$ is given by:
\begin{eqnarray*}
\ltir \ [L_\odot]  \sim  4.93 \times 10^{-22} (13.48{L}_\nu (12\mu\mbox{m}) + 5.16{L}_\nu (25\mu\mbox{m}) \\ +2.58{L}_\nu (60\mu\mbox{m})+ {L}_\nu (100\mu\mbox{m})) 
\end{eqnarray*}
where ${L}_\nu$ [$\mbox{erg\,s}^{-1}\mbox{Hz}^{-1}$] is defined as the luminosity per unit frequency at a frequency $\nu = c/\lambda$ where $c$ is the speed of light. 

In this work, L$_{\rm TIR}$ was computed using the 24 $\upmu$m luminosity (L$_{24 \rm\upmu m}$) via an empirical relation of \cite{2009ApJ...692..556R}; \cite[see, Equation  (A6) in ][]{2009ApJ...692..556R}, which is given by Equation  \ref{eq:Ltir-net}. 

\begin{equation}
\rm log \ L_{\rm TIR}  = (1.445\pm0.155) + (0.945\pm0.016)\ log \ L_{24 \rm\upmu m}
\label{eq:Ltir-net}
\end{equation}

Furthermore, in order to be consistent throughout our calculations and comparisons, we also used other formulations of  \cite{2009ApJ...692..556R} to estimate the L$_{60 \rm\upmu m}$ and L$_{\rm FIR}$. These transformations are presented in the next sections.

\subsubsection{Inferring IR luminosity at 60 $\mu$m (L$_{60 \rm\upmu m}$) }
We computed the L$_{60 \rm\upmu m}$ using the following relation taken from \cite{2009ApJ...692..556R}; \cite[see, Equation  (A7) in ][]{2009ApJ...692..556R}, which is given by Equation  \ref{eq:L60-net}. 

\begin{equation*}
\rm log(L_{\rm TIR}) = (1.183\pm0.101) + (0.920\pm0.010) \ log(L_{60 \rm\upmu m});
\end{equation*}

\begin{equation}
\rm Hence, \ log(L_{60 \rm\upmu m}) = \frac{log(L_{\rm TIR}) - 1.183}{0.920} 
\label{eq:L60-net}
\end{equation}

\subsubsection{Inferring far-IR luminosity (L$_{\rm FIR}$)}
It is common to study the relationship between the IR and radio luminosity using the classical far-IR luminosity as defined by \cite{1988ApJS...68..151H} at $\lambda = 42\mbox{--}122\,\mu\mbox{m}$. The far-IR luminosity L(FIR; 42--122 $\upmu$m) estimator is given by:

\begin{equation*}
\lfir \ [L_\odot] \sim 3.29 \times 10^{-22} \times \left(2.58{L}_\nu(60\mu\mbox{m}) + {L}_\nu(100\mu\mbox{m}) \right) 
\end{equation*}
where ${L}_\nu$ [$\mbox{erg\,s}^{-1}\mbox{Hz}^{-1}$] is defined as the luminosity per unit frequency at a frequency $\nu = c/\lambda$ where $c$ is the speed of light.

In this work, we estimated the L$_{\rm FIR}$ based on the assumption that the global ratio of L$_{\rm TIR}$, L(TIR; 8--1000 $\upmu$m), and L$_{\rm FIR}$ L(FIR; 42--122 $\upmu$m) luminosity is approximately 2 \citep[see e.g.,][]{2003ApJ...586..794B}, and see also in L$_{\rm TIR}$ defined as L(TIR; 3--1100 $\upmu$m) \citep[see e.g.,][]{2001ApJ...549..215D,2002ApJ...576..159D}. We have adopted the following relation as shown in Equation \ref{eq:Lfir-net}.

\begin{equation}
\rm L_{\rm TIR}/L_{\rm FIR} \sim 2; \ or \ L_{\rm FIR} \sim 0.5 \times L_{\rm TIR}
\label{eq:Lfir-net}
\end{equation}

\begin{center}
\begin{table*}
\label{tab:A-B-values} 
\captionsetup{justification=centering}
\caption{SFR(flux) fit coefficients A and B as a function of redshift for MIPS. {$A(z)$ and $B(z)$ are to be used in Equation \ref{eq:L24net01} and \ref{eq:L24net02} as the intercept and slope of the relation of SFR on observed IR flux \citep[see,][]{2009ApJ...692..556R}.}}
\begin{tabular}{l l l} \hline
  $z$ & $A_{24}$ &  $B_{24}$  \\  \hline \hline
 0.0 &  0.417 &  1.032  \\ \hline
 0.2 &  0.502 &  1.169  \\ \hline
 0.4 &  0.528 &  1.272  \\ \hline
 0.6 &  0.573 &  1.270  \\ \hline
 0.8 &  0.445 &  1.381  \\ \hline
 1.0 &  0.358 &  1.565  \\ \hline
 1.2 &  0.505 &  1.745  \\ \hline
 1.4 &  0.623 &  1.845  \\ \hline
 1.6 &  0.391 &  1.716  \\ \hline
 1.8 &  0.072 &  1.642  \\ \hline
 2.0 &  0.013 &  1.639  \\ \hline
 2.2 &  0.029 &  1.646  \\ \hline
 2.4 &  0.053 &  1.684  \\ \hline
 2.6 &  0.162 &  1.738  \\ \hline
 2.8 &  0.281 &  1.768  \\ \hline
 3.0 &  0.371 &  1.782  \\ \hline
\end{tabular} 
\end{table*}
\end{center}

\begin{center}
\begin{table*}
 \caption{This table summarizes various range of the IR and radio luminosity ratios mean values found in the literature. A non-exhaustive list of paper is presented as per the following: reference, redshift, environment, value $q$$_{\rm 24}$, $q$$_{\rm FIR}$ and value of $q$$_{\rm TIR}$ along with the available either a dispersion or an error of the mean.}
  \begin{tabular}{l l l l l l} \hline
    Reference  & Redshift & Environment & Mean $q$$_{\rm 24}$ & Mean $q$$_{\rm FIR}$ & Mean $q$$_{\rm TIR}$ \\ 
    \hline \hline
     This work &  $\sim$0.54 & Cluster & 0.69$\pm$0.55 & 1.80$\pm$0.53; $\lambda$(42--122) & 2.10$\pm$0.53; $\lambda$(8--1000)  \\ \hline
    \cite{2004ApJ...600..695R} &  0.0120$<$z$<$0.025 & Cluster & --------  & 2.07$\pm$0.74; $\lambda$(42--122) & --------  \\ \hline
    \cite{2001ApJ...554..803Y} & $\leq$0.16 & Field &  -------- & 2.34$\pm$0.01; $\lambda$(42--122) &  -------- \\ \hline
    \cite{2009ApJ...694.1435M} & $\sim$0.0036 & Cluster &  --------  & 2.10$\pm$0.25; $\lambda$(42--122)   &  -------- \\ \hline
    \cite{2009ApJ...692..556R} & $\leq$0.088 & Field &  1.22$\pm$0.02 & 2.42$\pm$0.23; $\lambda$(42--122) & -------- \\ \hline
    \cite{2001AJ....121.1903M} & 0.016$<$z$<$0.033 & Cluster & --------  & 2.30$\pm$0.20; $\lambda$(42--122) &  -------- \\ \hline
    \cite{1995AJ....109.1582A} & $<<$0.2 & Cluster & --------  & 2.27$\pm$0.20; $\lambda$(42--122)  &  --------  \\ \hline
    
    \cite{2009MNRAS.394.1685Y} & 1.5$<$z$<$3.0 & Field &  --------  & 2.23$\pm$0.04; $\lambda$(40--120) &   --------  \\ \hline
    \cite{2008ApJ...683..659S} & 0.5$<$z$<$3.0  & Field & -------- & 2.07$\pm$0.01; $\lambda$(40--120) &  --------  \\ \hline
    \cite{2006ApJ...650..592K} & 1$<$z$<$3 & Field &  --------  & 2.07$\pm$0.09; $\lambda$(42--122)  &  --------  \\ \hline
    Garrett (2002) & $\leq$1.4 & Field &  -------- & 2.00; $\lambda$(40--120) &  -------- \\ \hline    
    \cite{1985ApJ...298L...7H} & $\sim$0.0036  & Field & --------  & 2.14$\pm$0.14; $\lambda$(42--122) & --------  \\ \hline
    
    \cite{2011MNRAS.410.1155B} & 0$<$z$<$2 & Field & 1.47$\pm$0.03 & -------- & 2.66$\pm$0.12; $\lambda$(8--1000) \\ \hline 
    \cite{2010ApJS..186..341S} & 0$<$z$<$5 & Field & 1.26$\pm$0.13  & -------- & 2.57$\pm$0.13; $\lambda$(8--1000) \\ \hline 
    \cite{2010ApJ...714L.190S} & 0$<$z$<$2 & Field & -------- & -------- & 2.585$\pm$0.245; $\lambda$(8--1000) \\ \hline
    \cite{2010MNRAS.402..245I} & 0$<$z$<$3 & Field & --------  & -------- & 2.41$\pm$0.20; $\lambda$(8--1000) \\ \hline
    Ivison et al. (2010b) & 0$<$z$<$2 & Field & --------  & -------- & 2.40$\pm$0.24; $\lambda$(8--1000) \\ \hline
    \cite{2010MNRAS.409...92J} & 0$<$z$<$0.5 & Field & --------   & --------  & 2.40$\pm$0.12; $\lambda$(8--1000) \\ \hline   
    \cite{2009ApJ...698.1380M} & 0.6$\leq$z$\leq$2.6 & Field &  --------   &  --------  &  2.41$\pm$0.30; $\lambda$(8--1000) \\ \hline
    \cite{2003ApJ...586..794B} & local & Field &  --------  & 2.36$\pm$0.02; $\lambda$(42--122) & 2.64$\pm$0.02; $\lambda$(8--1000) \\ \hline
    \cite{2009MNRAS.397.1101G} & 0$<$z$<$2 & Field & 0.92$\pm$0.10 &  --------  &   --------  \\ \hline                
    \cite{2008MNRAS.385.1143B} & 0$<$z$<$1.2 & Field & 0.52$\pm$0.20 &  --------  &   --------  \\ \hline
    \cite{2008MNRAS.386..953I} & $\leq$3.5 & Field & 0.71$\pm$0.47 & -------- &  -------- \\ \hline
    \cite{2007MNRAS.376.1182B} & $\leq$2.15 & Field & 1.39$\pm$0.02 &  -------- & -------- \\ \hline
    \cite{2006ApJ...638..157M} & $\leq$0.002 & Field &  0.92$\pm$0.35 & 2.33$\pm$0.14; $\lambda$(42--122)   &  --------  \\ \hline
    \cite{2004ApJS..154..147A} & $\leq$2 & Field & 0.94$\pm$0.23 & --------   &  --------  \\ \hline     
  \end{tabular}
  \label{tab:q-define}
\end{table*}
\end{center}


\label{lastpage}

\end{document}